\documentclass[fleqn,twoside,twocolumn,nofootinbib,showkeys]{revtex4} % Specifies the document class %,unsortedaddress
\usepackage[sec,nocpr]{ujp} % \usepackage[cyr]{ujp} for cyrillic, \usepackage[web]{ujp} for web
\begin{document}
\title[Supercritical Instability]%колонтитул
{SUPERCRITICAL INSTABILITY OF DIRAC ELECTRONS IN THE FIELD OF TWO
OPPOSITELY CHARGED NUCLEI}%назва
\author{O.O.~Sobol}%1 автор
\affiliation{Taras Shevchenko National University of Kyiv}%институт
\address{2, Academician Glushkov Ave., Kyiv 03022, Ukraine}%адрес
\email{sobololeks@gmail.com}%e-mail
 \udk{???} \pacs{27.90+b, 03.65.Pm,\\[-3pt] 31.30.Jc} \razd{\seci}

\autorcol{O.O.\hspace*{0.7mm}Sobol}

\setcounter{page}{759}%

\begin{abstract}
The Dirac equation for an electron in a finite dipole potential has
been studied within the method of linear combination of atomic
orbitals (LCAO).\,\,The Coulomb potential of the nuclei that compose
a dipole is regularized, by considering the finite nuclear
size.\,\,It is shown that if the dipole momentum reaches a certain
critical value, the novel type of supercriticality occurs; namely,
the wave function of the highest occupied electron bound state
changes its localization from the negatively charged nucleus to the
positively charged one.\,\,This phenomenon can be interpreted as a
spontaneous creation of an electron-positron pair in vacuum, with
each of the created particles being in the bound state with the
corresponding nucleus and partially screening~it.
\end{abstract}
\keywords{supercritical instability, wave-function localization
change, critical dipole moment, LCAO method.} \maketitle

\section{Introduction}

\label{introduction}

The spectrum of a hydrogen-like atom has always been one of the
first test that every new wave equation was undergone.\,\,The
equation written down by Dirac in 1928 was not an exception
\cite{Dirac}.\,\, In the same year, Darwin and Gordon \cite{Darwin,
Gordon} obtained energy levels for a one-electron atom with the
nuclear charge $Ze$:%
%1
\begin{equation}
E=mc^{2}\left[  1+\frac{(Z\alpha)^{2}}{(\sqrt{(j+1/2)^{2}-(Z\alpha)^{2}}%
+n_{r})^{2}}\right]  ^{-{1}/{2}},
\end{equation}
where $j$ is the quantum number that characterizes the total angular
momentum, and $n_{r}$ is the radial quantum number.\,\,For the
ground state, $j=1/2$ and $n_{r}=$ $=0$, so that
$E_{0}=mc^{2}\sqrt{1-(Z\alpha)^{2}}$.\,\,According to this formula,
the ground-state energy becomes imaginary for superlarge nuclear
charges, $Z>\frac{1}{\alpha}\approx137$.\,\,This phenomenon was
called the \textquotedblleft falling to the
\mbox{center}\textquotedblright.

In 1945, I.Ya.\,\,Pomeranchuk and Ya.A.~Smorodinsky showed
\cite{finite-size} that this falling to the center results from the
singularity of the exact Coulomb potential at the coordinate
origin.\,\,The solution of the relativistic Kepler problem with
regard for finite dimensions of an atomic nucleus demonstrated that
there are no features at $Z=137$: the ground energy level
monotonically falls down, crosses the zero, and, if the so-called
critical value of nuclear charge, $Z_{\mathrm{cr}}\approx170$, is
exceeded, plunges into the lower continuum, so that the system
becomes unstable with respect to the creation of electron-positron
pairs \cite{Greiner,Zeldovich}.\,\,The electrons fill the $K$-shell
and partially screen the nucleus charge, whereas the free positrons
escape to infinity.\,\,However, there are no nuclei with so large
charges; therefore, the effect was not observed.

Later, there emerged an idea concerning front (or almost front)
collisions between the nuclei of heavy atoms, for example, uranium
\cite{Gershtein,Rafelski,Muller,Zeldovich}.\,\,In this case, their
total charge exceeds the critical value, and there exists such a
distance between the nuclei, at which the lowest bound state
immerses into the lower continuum.\,\,This distance is also called
critical.\,\,Un\-for\-tu\-na\-tely, in the relativistic problem with
two centers, the variables cannot be separated in any coordinate
system, so that an analytic solution cannot be constructed
\cite{Greiner,Zeldovich}.\,\,However, the corresponding calculations
were performed with the use of approximate quantum-mechanical
methods, in particular, the variational one \cite{Popov}, and the
dependences of the critical distance between the nuclei on the total
charge of the system were obtained.

\vspace*{-0.5mm}Interest to the supercritical instability problem
considerably grew after the experimental discovery of graphene in
2004 \cite{Geim}.\,\,Really, the low-energy electron excitation in
this two-dimensional crystal have an ultrarelativistic dispersion
law (a Dirac cone), being described by a (2\,+\,1)-dimensional
massless Dirac equation \cite{Gusynin, Guinea, Abergel}.\,\,Using a
series of experimental techniques, it is possible to create a mass
gap in the graphene spectrum \cite{Guinea, Giovanetti, Ponomarenko}
and obtain a (2\,+\,1)-dimensional analog of quantum
electrodynamics.\,\,In this case, the Fermi velocity plays the role
of light speed, $v_{\rm F}\approx\frac{c}{300}$, so that the
coupling constant (an
analog of the fine-structure constant) is much larger, $\tilde{\alpha}%
=\frac{e^{2}}{\hbar v_{\rm
F}}\approx\frac{300}{137}\approx2.2$.\,\,Although this value becomes
appreciably smaller owing to the dielectric permittivity of the
substrate for graphene, the critical impurity charge, at which the
supercritical instability can be induced, remains comparable with
unity \cite{Gamayun}.\,\,Re\-cent\-ly, the phenomenon of
supercritical instability in clusters of charged Ca dimers was
experimentally found \cite{Crommie}.\,\,In works \cite{SGG1, SGG2},
the supercritical collapse in the simplest cluster composed of two
identical impurities was studied, and the dependence of the critical
distance between them on their charge and mass gap was
calculated.\,\,In the case of gapless graphene, the instability was
shown to arise, when the total charge of impurities exceeds the
critical value irrespective of the distance between the impurities.

\vspace*{-0.5mm}In work \cite{Egger}, a problem of two different
impurities in graphene with a gap was considered.\,\,Confining the
analysis to the point-like dipole case, the cited authors examined
the features in the discrete spectrum of an electron in a vicinity
of continua and revealed an exponential condensation of levels,
which is similar to the Efimov scaling.\,\,However, a conclusion was
drawn that there cannot be any supercritical instability in this
system.\,\,This problem was also considered in works
\cite{Gorbar-dipole1, Gorbar-dipole2}.\,\,The supercritical
instability was shown to take place in the case of finite dipole,
but it should manifest itself differently: as a migration of the
wave function of the highest filled electron state.\,\,This
phenomenon was interpreted as a creation of an electron-hole pair in
vacuum, with each of the created particles being bound with the
corresponding impurity and screening its charge.\,\,This research
stimulated the consideration of a similar problem in the
(3\,+\,1)-dimensional QED, which is carried out in the presented
work.

The non-relativistic problem with the Schr\"{o}dinger equation was
repeatedly considered in the literature \cite{Fermi, Turner,
Abramov, Connolly}.\,\,It was found that, in the three-dimensional
space, there is a critical value for the
dipole moment, below which no bound states can exist in the system:%
%2
\begin{equation}
D_{\mathrm{cr}}=0.6393\ ea_{0}, \label{criticaldipole}%
\end{equation}
where $a_{0}=\frac{\hbar^{2}}{me^{2}}=0.529~\mathrm{\mathring{A}}$
is the Bohr radius.\,\,In work \cite{Connolly}, it was also shown
that the corresponding critical values of dipole moment equal zero
in one- and two-dimensional spaces, i.e.\,\,an arbitrarily small
dipole moment would generate bound states.

The relativistic problem with the Dirac equation for an electron in
the dipole potential was considered earlier in work
\cite{Matveev}.\,\,The cited authors studied the energy behavior, as
well as the corresponding wave functions, in the regions near the
continua using the method of asymptotic matching.\,\,At large
distances from the dipole, the squared Dirac equation has the form
of Schr\"{o}dinger equation and allows an asymptotic separation of
variables, which makes it possible to determine the critical dipole
moment.\,\,The latter is approximately equal to non-relativistic
value (\ref{criticaldipole}).

The problem with the electric dipole is evidently symmetric with
respect to the charge inversion.\,\,This symmetry plays the crucial
role and gives rise to a symmetric arrangement of energy levels
relatively to zero.\,\,As a result, the electron and positron
(symmetric to it) states could have intersected only at $E=0$.
However, this is impossible, because those states have identical
quantum numbers, and the Wigner--von Neumann theorem on the absence
of a crossing of levels is applicable \cite{Wigner}.\,\,This
circumstance brings about the
existence of characteristic \textquotedblleft constrictions\textquotedblright%
\ in the spectrum (cf.\,\,Figs.~5 and 7 below).

The consideration of the Dirac equation with an electric dipole
field has also a relation to the description of the heavy meson
decay \cite{Greiner}.\,\,Really, a heavy quark and an antiquark have
opposite color charges and, therefore, form a dipole.\,\,In the
color field of this dipole (which, owing to the confinement
phenomenon, increases when the quarks are attempted to move some
distance apart), a pair of a light quark and an antiquark can be
created.\,\,This pair gets bound with a heavy quark and an antiquark
and screens their charges.\,\,As a result, the decay of one heavy
meson gives rise to the emergence of two lighter ones.

In this work, the method of linear combination of atomic orbitals
(LCAO) will be applied to calculate the spectrum and the wave
functions of an electron in the potential created by a finite
electric dipole.\,\,The Coulomb potential of the nuclei that form
the dipole was regularized, by taking their finite dimensions into
account.\,\,The wave functions of several first electron levels in
the field of one nucleus, which are centered at the corresponding
nucleus, were used as atomic orbitals.\,\,Attempts were made to
extend the limits of applicability of the LCAO method to
supercritical nuclear charges.\,\,The corresponding calculations
were carried out not only near the edges of continua, but over the
whole mass gap, which has not been studied earlier in the
literature.

The paper is organized as follows.\,\,In Section \ref{Onenucleus},
an exact solution of the problem with a regularized potential of one
nucleus is quoted, and various types of regularizations are
analyzed.\,\,In Section \ref{LCAO}, within the LCAO method, the
energy and wave functions of an electron in the dipole potential are
calculated, and their behavior depending on the dipole moment
variation is considered.\,\,Section \ref{extended_LCAO} is devoted
to the extension of the limits of applicability of the LCAO method
to nuclei with large charge values.\,\,The obtained results are
generalized and the conclusions are drawn in Section
\ref{conclusions}.\,\,Some technical calculations that arise in the
course of the LCAO method application are presented in Appendix.

\section{Regularized Potential of a Single Nucleus}

\label{Onenucleus}

Let us consider the motion of an electron in the potential of a
charged nucleus.\,\,The corresponding Hamiltonian looks like
%3
\begin{equation}
\hat{H}=-i\hbar c
\gamma^{0}{\boldsymbol\gamma}{\boldsymbol\nabla}+\gamma^{0}mc^{2}-Ze^{2}v(r),
\end{equation}
where the function $v(r)$ describes the regularized Coulomb
potential, and $\gamma^{\mu}$ ($\mu=0,...,3$) are the Dirac
matrices, for which the standard Dirac--Pauli representation is
used.

To make calculations convenient, let us change to dimensionless
variables.\,\,In what follows, all energy quantities will be
measured in terms of electron rest mass units,
$E_{0}=mc^{2}\approx0.511~\mathrm{MeV}$, making no changes of their
notations.\,\,All distances will be measured in terms of the Compton
wavelength for the electron, $\lambdabar_{C}=\frac{\hbar}{mc}\approx
386~\mathrm{fm}$.\,\,We also introduce the notation $\zeta=Z\alpha$,
where $\alpha=\frac{e^{2}}{\hbar c}\approx\frac{1}{137}$ is the
fine-structure constant.

The potential of a charged nucleus is spherically symmetric, so that
the total angular momentum remains constant.\,\,The wave functions,
which are characteristic of the operators $\hat{J}^{2}$ and
$\hat{J}_{z}$, should be sought in the form
%4
\begin{equation}
\label{wave-func} |\Psi\rangle=\left(\!\!
\begin{array}{c}
g(r)\Omega_{j,j_{z},l}\\[2mm]
if(r)\Omega'_{j,j_{z},l'}
\end{array}
\!\!\!\right)\!\!.
\end{equation}
Here, $l=j-1/2$, $l^{\prime}=l+1=2j-l$, the wave function parity
equals $(-1)^{l}$, and the angular dependences are described by the
spherical spinors
$\Omega$ and $\Omega^{\prime}$, which look like%
%5
\begin{equation}
\begin{array}{l}
\Omega_{j,j_{z},l}=\left(\!\!
\begin{array}{c}
\sqrt{\frac{j+j_{z}}{2j}}Y_{l,j_{z}-1/2}\\[2mm]
\sqrt{\frac{j-j_{z}}{2j}}Y_{l,j_{z}+1/2}
\end{array}
\!\!\right)\!\!,\\[6mm]
\Omega'_{j,j_{z},l'}=\left(\!\!
\begin{array}{c}
-\sqrt{\frac{j-j_{z}+1}{2j+2}}Y_{l',j_{z}-1/2}\\[2mm]
\sqrt{\frac{j+j_{z}+1}{2j+2}}Y_{l',j_{z}+1/2}
\end{array}
\!\!\right)\!\!,\end{array}
\end{equation}
where the spherical harmonics are designated in the standard way
\cite{Landau},
%6
\[
Y_{l,m}(\theta,\varphi)=(-1)^{\frac{m+|m|}{2}}i^{l}\sqrt{\frac{2l+1}{4\pi}\frac{(l-|m|)!}{(l+|m|)!}}\,\times\]\vspace*{-7mm}
\begin{equation}
\times\, P^{|m|}_{l}(\cos\theta)e^{im\varphi}.
\end{equation}
The wave functions that have, for the same $j$-value, the opposite
parity in comparison with the functions indicated in
Eq.~(\ref{wave-func}) are tried in a similar form, in which the
spherical spinors $\Omega$ and $\Omega^{\prime}$ should be swapped,
%7
\begin{equation}
\label{wave-func-2} |\tilde{\Psi}\rangle=\left(\!\!
\begin{array}{c}
i\tilde{g}(r)\Omega'_{j,j_{z},l'}\\[2mm]
-\tilde{f}(r)\Omega_{j,j_{z},l}
\end{array}
\!\!\right)\!\!,
\end{equation}
where the additional phase multiplier $i$ was introduced for the
sake of convenience.

After substituting expressions~(\ref{wave-func}) and (\ref{wave-func-2}) into
the Dirac equation $\hat{H}|\Psi\rangle=E|\Psi\rangle$, the following systems
of ordinary differential equations for the radial functions are obtained:%
%8-9
\begin{eqnarray}
\label{system-eq} \hspace{-4mm}&&\left\{\!\!
\begin{array}{l}
g'-\frac{j-1/2}{r}g-f(1+E+\zeta v(r))=0,\\[2mm]
f'+\frac{j+3/2}{r}f-g(1-E-\zeta v(r))=0,
\end{array}
\right.\\
\label{system-eq-2} \hspace{-4mm}&&\left\{\!\!
\begin{array}{l}
\tilde{g}'+\frac{j+3/2}{r}\tilde{g}-\tilde{f}(1+E+\zeta
v(r))=0,\\[2mm]
\tilde{f}'-\frac{j-1/2}{r}\tilde{f}-\tilde{g}(1-E-\zeta v(r))=0.
\end{array}
\right.
\end{eqnarray}
The normalization condition for the wave functions, which involves
the orthonormal character of spherical spinors, acquires the form%
%10
\begin{equation}
\begin{array}{l}
\displaystyle\langle {\Psi} |{\Psi}\rangle=\int\limits_{0}^{+\infty}r^{2}(f^{2}(r)+g^{2}(r))dr=1,\\[2mm]
\displaystyle\langle\tilde{\Psi}|\tilde{\Psi}\rangle=\int\limits_{0}^{+\infty}r^{2}(\tilde{f}^{2}(r)+\tilde{g}^{2}(r))dr=1.
\end{array}
\label{orthonorm}
\end{equation}

For every fixed value of quantum number $j$, the systems of
equations (\ref{system-eq}) and (\ref{system-eq-2}) have infinite
numbers of solutions that satisfy the normalization conditions
(\ref{orthonorm}).\,\,Therefore, let us mark those solutions by the
radial quantum number $n_{r}=0,1,2,...$\,.\,\,Furthermore, let us
designate the orbital quantum number $L$ of the spherical spinor
that appears in the upper component of the wave function by capital
Latin letters following the conventional order: $S\,(L=0)$,
$P\,(L=1)$, $D\,(L=2)$, $F\,(L=3)$, and so on.\,\,By analogy with
the non-relativistic case, let us also introduce the principal
quantum number $n=n_{r}+L+1$.\,\,The Russell--Saunders notation
\cite{Bethe} will be used to designate terms, and, additionally, the
projection of the angular momentum will be indicated: $|n\,L_{j},\
j_{z}\rangle$.\,\,Hence, the first (by energy) three states have the
following sets of quantum numbers:
%11
\begin{subequations}
\label{first-three-levels}%
\begin{align}
&1)~j=1/2,\ j_{z}=\pm1/2,\ n_{r}=0,\ L=l=0,\ l^{\prime}=1,\nonumber\\
&  n=1\ \rightarrow\ |1S_{1/2},\ \pm1/2\rangle;\\
&2)~j=1/2,\ j_{z}=\pm1/2,\ n_{r}=0,\ L=l^{\prime}=1,\ l=0,\nonumber\\
&  n=2\ \rightarrow\ |2P_{1/2},\ \pm1/2\rangle;\\
&3)~j=1/2,\ j_{z}=\pm1/2,\ n_{r}=1,\ L=l=0,\ l^{\prime}=1,\nonumber\\
&  n=2\ \rightarrow\ |2S_{1/2},\ \pm1/2\rangle.
\end{align}
\end{subequations}

The problem with the regularized Coulomb potential has no analytic
solution. Therefore, the systems of equations (\ref{system-eq}) and
(\ref{system-eq-2}) have to be integrated numerically.\,\,For this
purpose, we should determine boundary conditions for the radial
functions at $r=0$.\,\,Let us express $f$ (or $\tilde{f}$) from the
first equation of the system and substitute it into the second
one.\,\,Let the characteristic regularization scale be equal to
$r_{0}$. Then, if $r<r_{0}$, we obtain $v(r)\approx
v(0)\gg1$.\,\,The asymptotics can be
determined from the approximate equation%
%12
\begin{equation}
g''+\frac{2}{r}g'+\left(\!\zeta^{2}v^{2}(0)-\frac{j^{2}-1/4}{r^{2}}\!\right)g=0,
\end{equation}
Its solution that is regular at $r=0$ looks like%
%13
\begin{equation}
g(r)\sim\frac{1}{\sqrt{r}}J_{j}(\zeta v(0)r).
\end{equation}

The asymptotic behavior of the solutions of system
(\ref{system-eq-2}) is analyzed analogously.\,\,At $j=1/2$, the
boundary conditions can be chosen in the form
%14
\begin{equation}
\begin{array}{l}
g(0)=1,\ \ f(0)=0;\\[2mm]
\tilde{g}(0)=0,\ \ \tilde{f}(0)=1.
\end{array}
\end{equation}
The spectrum is determined from the condition that the wave
function have to fall down exponentially as $r\rightarrow\infty$.

The result depends on the specific manner of the potential
regularization.\,\,The following three variants should be
considered:

(i)~regularization by \textquotedblleft going into the fourth
dimension\textquotedblright:%
%15
\begin{equation}
v_{I}(r,r_{0})=\frac{1}{\sqrt{r^{2}+r_{0}^{2}}};
\end{equation}

(ii)~regularization by considering the finite nuclear size
(the nuclear charge is uniformly distributed over the nuclear surface):%
%16
\begin{equation}
v_{\rm II}(r,r_{0})=\left\{\!\!
\begin{array}{ll}
\displaystyle{\frac{1}{r}}, & r\geq r_{0};\\[3mm]
\displaystyle{\frac{1}{r_{0}}}, & r<r_{0}.
\end{array}
\right.
\end{equation}

(iii)~regularization by considering the finite nuclear size
(the nuclear charge is uniformly distributed over the nuclear volume):%
%17
\begin{equation}
v_{\rm III}(r,r_{0})=\left\{\!\!
\begin{array}{ll}
\displaystyle {\frac{1}{r}}, & r\geq r_{0};\\[3mm]
\displaystyle {\frac{1}{r_{0}}\frac{3-(r/r_{0})^{2}}{2}}, & r<r_{0}.
\end{array}
\right.
\end{equation}

To model the real value of regularizing parameter $r_{0}$, let us use the
empirical relation for the nuclear radius, which is known in nuclear physics
\cite{Marmier}:
%18
\begin{equation}
r_{0}\approx1.25\,\mathrm{fm}\cdot A^{1/3}, \label{r0_on_A}%
\end{equation}
where $A=Z+N$ is the nuclear mass number.\,\,In order to find the
dependence of nuclear radius on the nuclear charge, let us adopt
that the nucleus lies in the so-called beta-stability valley, for
which the following approximate relation is obeyed \cite{Marmier}:
%19
\begin{equation}
Z=\frac{A}{1.98+0.015A^{2/3}}.
\end{equation}
By solving the cubic equation, we obtain the dependence
$A=A(Z)$.\,\,Afterward, from Eq.~(\ref{r0_on_A}), we calculate the
dependence $r_{0}=r_{0}(Z)$.\,\,The latter is shown in Fig.~1.

Figure~2 demonstrates the dependences of the ground-state energy of
the electron in the field of nucleus on the nuclear charge for three
kinds of regularization.\,\,For comparison, a similar curve for the
case of non-regularized potential $E_{NR}=\sqrt{1-(Z\alpha)^{2}}$ is
also depicted.\,\,From the presented curves, one can see that the
\textquotedblleft falling to center\textquotedblright\ phenomenon
takes place for the non-regularized potential, if $Z\gtrsim137$,
i.e.\,\,when the ground-state energy becomes imaginary, and the
system has no definite ground state.\,\,This problem is eliminated,
if the potential is regularized in any fashion.\,\,The ground-state
energy monotonically falls down and becomes negative at a certain
threshold charge value.\,\,Depending on the regularization type, the
threshold charge value falls within the interval $148\leq
Z_{0}\leq152$.\,\,Then the energy continues to decrease
monotonically and, when achieving the critical value, the level
plunges into the lower continuum, inducing the supercritical
instability.\,\,Depending on the regularization type, the critical
charge falls within the interval $170\leq Z_{\mathrm{cr}}\leq186$.

The most natural is the regularization of the third kind (i.e.\,\,a
uniform charge distribution in a nucleus with finite
dimensions).\,\,Therefore, it will be used below in our all
numerical calculations, although, for the sake of generality, an
arbitrarily regularized potential $v(r)$ will be retained in all
relevant formulas.

%Рис 1
\begin{figure}
\vskip1mm
  \includegraphics[width=7.5cm]{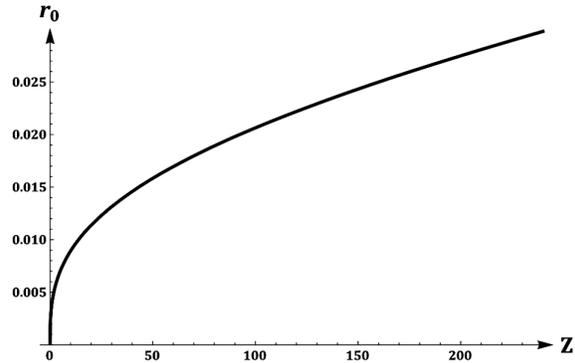}
  \vskip-3mm\caption{Dependence of the regularizing parameter $r_{0}$
on the nuclear charge $Z$}
  \label{Regularization}
\end{figure}
%Рис.\,\,2
\begin{figure}
\vskip1mm
  \includegraphics[width=7.5cm]{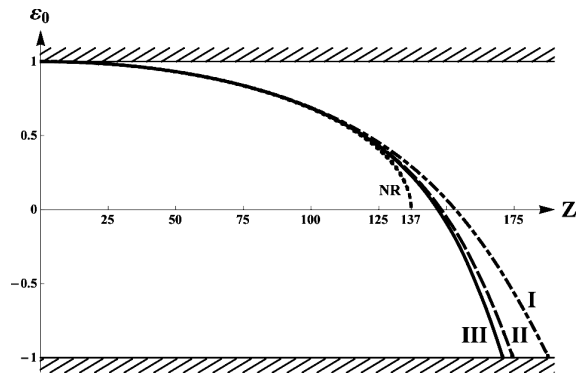}
  \vskip-3mm\caption{Dependences of the ground-state energy of
an electron in the regularized Coulomb potential on the nuclear
charge for various regularization models: I (dash-dotted curve), II
(dashed curve), and III (solid curve).\,\,For comparison, the same
dependence for the non-regularized (NR) potential is shown (dotted
curve)}
  \label{SpExact}
\end{figure}
%Рис.\,\,3
\begin{figure}
\vskip1mm
  \includegraphics[width=7.5cm]{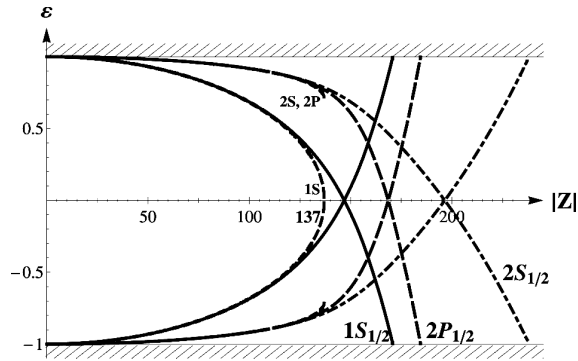}
  \vskip-3mm\caption{Energy dependences on the nuclear charge for
three lowest electron levels (\ref{first-three-levels}) in the
regularized Coulomb potential: $1S_{1/2}$ (solid curves), $2P_{1/2}$
(dashed curves), and $2S_{1/2}$ (dash-dotted curves).\,\,Red dashed
curves exhibit the energy of corresponding levels in the
non-regularized Coulomb potential}
  \label{FirstThreeLevels}
\end{figure}

%Fig.\,\,4
\begin{figure}
\vskip1mm
  \includegraphics[width=\column]{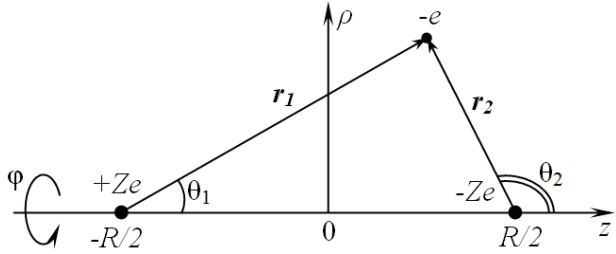}
  \vskip-3mm\caption{Schematic diagram of an electron in the field of
an electric dipole}
  \label{Geometry}
\end{figure}

In Fig.~3, the energy dependences on the absolute value of nuclear
charge are exhibited for three first levels
(\ref{first-three-levels}) of an electron in the regularized Coulomb
potential (regularization of type iii).\,\,The levels start from the
upper continuum for a positively charged nucleus and, symmetrically,
from the lower continuum for a negatively charged one.\,\,On the
basis of those three energy levels, the LCAO method for finding the
energy and wave functions of an electron in the dipole potential
will be developed in the next \mbox{sections.}\looseness=1

\section{LCAO Method\\ for the Problem with a Dipole}

\label{LCAO}

Now let us consider an electron in the potential of a finite
electric dipole.\,\,The corresponding dimensionless Hamiltonian
looks like:
%20
\begin{equation}
\label{hamiltonian-dipole}
\hat{H}=-i\gamma^{0}{\boldsymbol\gamma}{\boldsymbol\nabla}+\gamma^{0}-\zeta
\left(v(r_{1})-v(r_{2})\right),
\end{equation}
where $r_{1,2}=\sqrt{(z\pm R/2)^{2}+\rho^{2}}$ and
$\rho=\sqrt{x^{2}+y^{2}}$ (see Fig.~4).\,\,Unfortunately, none of
coordinate systems allow the variables in this problem to be
separated.\,\,Therefore, approximate methods will be applied.\,\,The
arrangement of the system in the space, coordinate system, and some
notations are shown in Fig.~4.

Let us perform the analysis, by using the LCAO method.\,\,This
technique is well-known and widely applied in molecular physics
\cite{Diu}.\,\,In this method, the wave functions are selected as
linear combinations of basis functions.\,\,The latter, as a rule,
are one-electron functions localized at the corresponding atoms in
the molecule concerned.\,\,The coefficients in a linear combination
are determined by minimizing the total energy of the system.

The Hamiltonians for an electron in the fields of positively and negatively
charged nuclei will be denoted as, respectively,%
%21
\begin{equation}
\hat{H}_{\pm}(\zeta)=-i\gamma^{0}{\boldsymbol\gamma}{\boldsymbol\nabla}+\gamma^{0}\pm\zeta
v(r).
\end{equation}
The charge conjugation operator
%22
\begin{equation}
\hat{U}_{c}=-i\gamma^{2}\hat{K},
\end{equation}
where $K$ is the complex conjugation operator, and $\hat{U}_{c}^{2}=1$,
transforms those Hamiltonians into each other:%
%23
\begin{equation}
\hat{U}_{c}\hat{H}_{+}(\zeta)\hat{U}_{c}=-\hat{H}_{-}(\zeta).
\end{equation}
Therefore, if $|\Psi\rangle$ is a characteristic function for the Hamiltonian
$H_{-}(\zeta)$ with energy $E$, the function $|\Psi_{c}\rangle=\hat{U}%
_{c}|\Psi\rangle$ is a characteristic function for the Hamiltonian
$\hat{H}_{+}(\zeta)$ with the energy $-E$.

Using the explicit expressions for the spherical spinors, it is easy to get
convinced that%
%24
\begin{equation}
\begin{array}{l}
\sigma_{2}\Omega^{*}_{j,j_{z},l}=-i(-1)^{(l+1)-j_{z}+1/2}\Omega_{j,-j_{z},l},\\[2mm]
\sigma_{2}\Omega'^{*}_{j,j_{z},l'}=-i(-1)^{l'-j_{z}+1/2}\Omega'_{j,-j_{z},l'},
\end{array}
\end{equation}
i.e. the charge conjugation operator changes the sign of the angular
momentum projection.

The problem with the dipole potential is no more spherically
symmetric.\,\,Therefore, the total momentum is not
conserved.\,\,However, the axial symmetry with respect to the axis
$Oz$ that passes through both charged centers still exists, so that
the projection of the total momentum on this axis is
constant.\,\,Let us choose a test function for the LCAO method as a
linear combination of wave functions with identical values of the
angular momentum projection,
$j_{z}=+1/2$:%
%25
\begin{equation}
|\Psi\rangle=c_{1}|1\rangle+c_{2}|2\rangle,\label{linear-combination}%
\end{equation}
where the basis states are%
%26
\[
|1\rangle=|1S_{1/2},  +1/2,  r_{1}, \zeta\rangle=\]\vspace*{-7mm}
\begin{equation}
 =\left(\!\!
\begin{array}{c}
g_{0}(r_{1})\Omega_{1/2,1/2,0}(1)\\[2mm]
if_{0}(r_{1})\Omega'_{1/2,1/2,1}(1)
\end{array}
\!\!\right)\!\!,
\end{equation}\vspace*{-5mm}
%27
\[|2\rangle=\hat{U}_{c}|1S_{1/2},\  -1/2,\  r_{2},\
\zeta\rangle=\]\vspace*{-7mm}
\begin{equation}
=\left(\!\!
\begin{array}{c}
if_{0}(r_{2})\Omega'_{1/2,1/2,1}(2)\\[2mm]
g_{0}(r_{2})\Omega_{1/2,1/2,0}(2)
\end{array}
\!\!\right)\!\!.
\end{equation}
In these expressions, the radial functions $f_{0}(r)$ and $g_{0}(r)$ are
determined by numerically integrating the problem with one nucleus (see
Section~\ref{Onenucleus}).\,\,The spherical spinors look like%
%28
\begin{equation}
\begin{array}{l}
\Omega_{1/2,1/2,0}=\displaystyle{\frac{1}{\sqrt{4\pi}}}\left(\!
\begin{array}{c}
1\\
0
\end{array}
\!\right)\!\!,\\
\Omega'_{1/2,1/2,1}=\displaystyle{\frac{i}{\sqrt{4\pi}}}\left(\!\!
\begin{array}{c}
\cos\theta\\
-\sin\theta e^{i\varphi}
\end{array}
\!\!\right)\!\!.
\end{array}
\end{equation}

Let us project the Dirac equation $\hat{H}|\Psi\rangle=E|\Psi\rangle$ on each
of the basis states:%
%29
\begin{equation}
\begin{array}{l}
c_{1}H_{11}+c_{2}H_{12}=E(c_{1}+c_{2}S),\\[2mm]
c_{1}H_{21}+c_{2}H_{22}=E(c_{1}S+c_{2}).
\end{array}
\end{equation}
Then we obtain the following secular equation for the non-trivial
coefficients in
the linear combination (\ref{linear-combination}):%
%30
\begin{equation}
\det \left|\!
\begin{array}{cc}
H_{11}-E & H_{12}-SE\\[2mm]
H_{21}-S^{*}E & H_{22}-E
\end{array}
\!\right|=0,
\end{equation}
where $H_{ij}=\langle i|\hat{H}|j\rangle$ are the matrix elements of
the Hamiltonian, and $S=\langle1|2\rangle$ is the overlap integral,
which equals zero as is shown by formula (\ref{Sint}) in
\mbox{Appendix.}

In order to calculate the matrix elements, we take the total Hamiltonian in
the form
\[
\hat{H}=\hat{H}_{-}(1)+\zeta v(r_{2})=\hat{H}_{+}(2)-\zeta v(r_{1}).
\]
Then,
%31,a,b
\begin{subequations}
\begin{eqnarray}
\hspace{-4mm}&&H_{11}=-H_{22}=\varepsilon_{0}+\zeta C,\\
\hspace{-4mm}&&H_{12}=H_{21}=-\zeta A,
\end{eqnarray}
\end{subequations}
where $\varepsilon_{0}=\varepsilon^{(1S)}(\zeta)$ is the
ground-state energy of an electron in the field of one nucleus (see
Figs.~2 and 3).\,\,The expressions for the matrix elements include
the Coulomb, $C$, and exchange, $A$, integrals.\,\,Expressions for
them can be found in Appendix: formulas (\ref{Coul}) and
(\ref{Exchange}), respectively. These integrals are calculated
provided that the functions $f_{0}$ and $g_{0}$ are given.

Ultimately, we obtain a spectrum%
%32
\begin{equation}
\label{spectrum}
\varepsilon=\pm\sqrt{(H_{11})^{2}+(H_{12})^{2}}=\pm\sqrt{(\varepsilon_{0}+\zeta
C)^{2}+\zeta^{2}A^{2}},
\end{equation}
which is symmetric with respect to zero and reflects the charge
symmetry of the problem.\,\,For the negative spectral branch, the
coefficients in the linear
combination (\ref{linear-combination}) are as follows:%
%33
\begin{equation}
\begin{array}{l}
\displaystyle c_1=\frac{H_{12}}{\sqrt{H_{12}^{2}+(H_{11}+\sqrt{H_{11}^{2}+H_{12}^{2}})^{2}}},\\[4mm]
\displaystyle
c_{2}=-\frac{H_{11}+\sqrt{H_{11}^{2}+H_{12}^{2}}}{\sqrt{H_{12}^{2}+(H_{11}+\sqrt{H_{11}^{2}+H_{12}^{2}})^{2}}}.
\end{array}
\end{equation}
It is easy to see that, as $R\rightarrow\infty$, we have $C\approx\frac{1}%
{R}\rightarrow0$ and
$A\sim\exp(-\sqrt{1-\varepsilon_{0}^{2}}R)\rightarrow0$.\,\,Therefore,
at large distances between the nuclei, the energy of the system
$\varepsilon\rightarrow\pm\left\vert \varepsilon_{0}\right\vert
$.\,\,This is quite expectedly, since we may consider the nuclei to
be isolated in this case.

From expression (\ref{spectrum}) for the energy and from the fact
that $C\geq0$, one can see that the spectral \textquotedblleft
constriction\textquotedblright\ can arise only, if
$\varepsilon_{0}<0$, with the minimum itself being located near the
point, at which $\zeta C=\vert \varepsilon_{0}\vert $. At the same
point, we have $\vert c_{1}\vert =\vert c_{2}\vert =1/\sqrt{2}$,
i.e.\,\,the wave
function changes its localization:%
%34
\[
|c_{2}|^{2}-|c_{1}|^{2}=\frac{2H_{11}(H_{11}+\sqrt{H_{11}^{2}+H_{12}^{2}})}{H_{12}^{2}+(H_{11}+\sqrt{H_{11}^{2}+H_{12}^{2}})^{2}}\sim\]
\vspace*{-7mm}
\begin{equation}
 \sim{\rm sign}(H_{11}).
\end{equation}
The quantities $C$ and $\vert A\vert $ monotonically decrease with
the growth of the distance $R$.\,\,Initially, $\vert c_{2}\vert
>\vert c_{1}\vert $, i.e.\,\,the electron charge density is mainly
localized at the negatively charged nucleus.\,\,After passing the
\textquotedblleft const\-ric\-tion\textquotedblright, we obtain
$\vert c_{2}\vert <\vert c_{1}\vert $, i.e.\,\,the wave function
changes its localization to the positively charged nucleus.

One can also see that if $\varepsilon_{0}>0$, we always obtain
$H_{11}>0$ and $\left\vert c_{2}\right\vert >\left\vert
c_{1}\right\vert $.\,\,As a result, the wave function remains mainly
localized at the negatively charged nucleus, i.e.\,\,its
localization does not change.

%Fig.\,\,5
\begin{figure}
\vskip1mm
  \includegraphics[width=8cm]{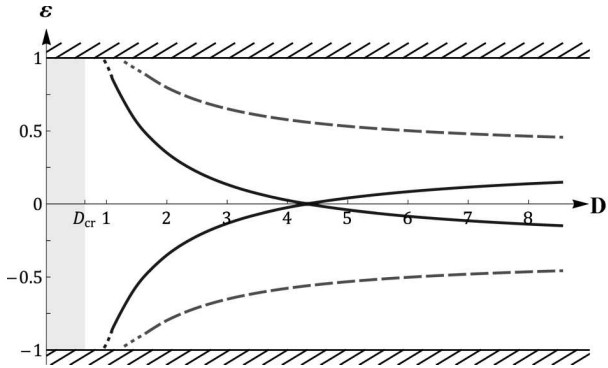}
\vskip-3mm\caption{Dependences of the ground-state energy of
an electron in the field of
an electric dipole on the dimensionless dipole moment $D=\frac{D_{\rm ph}}{ea_{0}%
}=\zeta R$ for two values of nuclear charge: $Z=130$ (dashed curves)
and 156 (solid curves).\,\,The tinted area marks the region with no
bound states [see formula (\ref{criticaldipole})]}
  \label{LCAO_energy}
\end{figure}

%Fig.\,\,6
\begin{figure}
\vskip1mm
  \includegraphics[width=8cm]{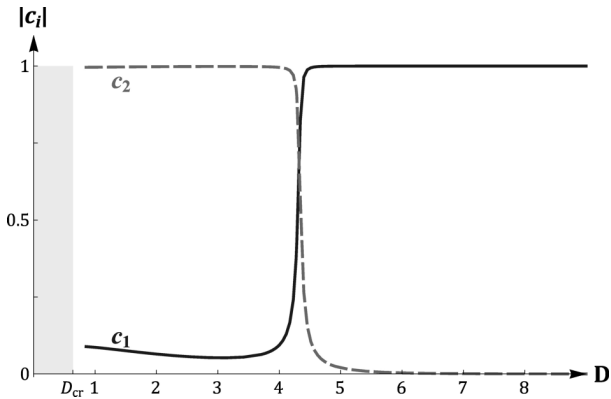}
\vskip-3mm\caption{Dependences of the coefficients in the linear
combination (\ref{linear-combination}) for the highest filled
electron state (the lower
solid curve in Fig.~5) on the dimensionless dipole moment $D=\frac{D_{\rm ph}%
}{ea_{0}}=\zeta R$ for the nuclear charge $Z=156$.\,\,The phenomenon
of wave-function localization change can be observed}
  \label{LCAO_coef}
\end{figure}

Figure 5 exhibits energy curves for the ground state of an electron
in the field of an electric dipole that were calculated for two
values of nuclear charge.\,\,At $\zeta=0.95$ ($Z=130$), the
ground-state energy of an electron in the field of one nucleus is
positive ($\varepsilon_{0}=+0.35$, see Fig.~3); therefore, the
electron energy monotonically changes in the dipole field as the
dipole moment grows (dashed curves in Fig.~5).\,\,At $\zeta=1.14$
($Z=156$), the ground state of an electron in the field of one
nucleus is negative ($\varepsilon_{0}=-0.30$, see Fig.~3);
therefore, the electron energy in the dipole field passes a
\textquotedblleft constriction\textquotedblright\ (solid curves in
Fig.~5), and the wave function of the highest filled state changes
its localization at this moment from the negatively charged nucleus
to the positively charged one (Fig.~6).

All levels with negative energies start from the lower continuum,
when the dipole moment exceeds some critical value
(\ref{criticaldipole}).\,\,All of them are filled with electrons
from the Dirac sea.\,\,Those \textquotedblleft
sea\textquotedblright\ electrons are initially localized at the
negatively charged nucleus.\,\,Then, as the distance between the
nuclei increases (the dipole moment grows), the wave function of one
of those sea electrons migrates to the positively charged
nucleus.\,\,A hole (positron) remains at its place near the
negatively charged nucleus.\,\,In such a manner, we obtain the
electron-positron pair created from vacuum.\,\,Each of the particles
is in the bound state with the corresponding nucleus and partially
screens its charge.\,\,It should be noted that, for this phenomenon
to emerge, the nuclear charges must be so large that the level of an
electron in the field of a positive nucleus would cross the energy
zero, or the level of an electron in the field of a positive nucleus
and its symmetric counterpart in the field of a negative nucleus
would together overcome the energy gap equal to $2mc^{2}$.

\vspace*{-0.5mm}The described phenomenon is very similar to the
supercritical atomic collapse.\,\,The difference consists in that,
in the collapse case, we have one nucleus with the charge
$Z\gtrsim170$, and, in its field, the lowest electron bound state
can reach the lower continuum on its own (in so doing, it crosses
the energy distance $2mc^{2}$).\,\,The system becomes unstable with
respect to the creation of electron-positron pairs.\,\,The created
electrons are localized at the nucleus and partially screen its
charge, whereas the free positrons escape to infinity.

\vspace*{-0.5mm}In the framework of this consideration, a
shortcoming of the LCAO method is the fact that the asymptotic
behavior of the wave functions at infinity is $\sim$$
e^{-\sqrt{1-\varepsilon_{0}^{2}}r}$ rather than $\sim$$ e^{-\sqrt
{1-\varepsilon^{2}}r}$, as it should be.\,\,This shortcoming is
especially pronounced at short distances between the nuclei, $R$,
when the true energy of the system tends to the continuum
boundaries, so that the exponential damping has to be weak.\,\,In
this method, however, the wave functions are constructed from the
wave functions obtained in the problem with one
center.\,\,Therefore, the character of their damping is identical at
any distance, being only determined by the ground state energy of an
electron in the field of one center.\,\,Hence, the method concerned
is inapplicable at small $R$.

The described method is based only on two terms in the linear
combination (\ref{linear-combination}).\,\,In other words, it takes
into account only the ground state of an electron in the field of
one nucleus.\,\,This scenario is rather good for relatively small
charges.\,\,However, if the charge of nuclei is so large that the
ground level in the field of a positive nucleus and the second level
in the field of a negative nucleus have approximately the same
energy ($Z\gtrsim160$, see Fig.~3), the linear combination has also
to include the wave functions of the second energy level.\,\,The
situation becomes even more worsened for supercritical values of
nuclear charges, when the ground level plunges into the lower
continuum ($Z\gtrsim170$, see Fig.~3).\,\,In this case, the
ground-state energy becomes complex, and the wave functions
non-normalized, so that the application of the latter as basis
functions becomes impossible.\,\,These difficulties can be resolved
by making a certain modification to the LCAO method, which is
described in the next section.

\section{Extension of Applicability\\ Range of LCAO Method}

\label{extended_LCAO}

In order to overcome the difficulties described above, let us try the wave
function as the following linear combination (the indicated number of terms
is enough for nuclei with the charges of nuclei $Z\lesssim185$; for larger
charges, the next excited states should be taken into consideration):%
%35
\begin{equation}
|\Psi\rangle=\sum\limits_{i=1}^{6}c_{i}|i\rangle, \label{linear-combination2}%
\end{equation}
where the states $|i\rangle$ look like:
%36a
\begin{subequations}
\[
  |1\rangle=|1S_{1/2}, +1/2,\ r_{1}, \zeta_{a}%
\rangle=\]\vspace*{-9mm}
\begin{equation}
  =\left(\!\!
\begin{array}
[c]{c}%
g_{0}(r_{1})\Omega_{1/2,1/2,0}(1)\\[2mm]
if_{0}(r_{1})\Omega_{1/2,1/2,1}^{\prime}(1)
\end{array}
\!\!\right)\!\!  ,
\end{equation}\vspace*{-5mm}
%36б
\[
|2\rangle=\hat{U}_{c}|1S_{1/2}, -1/2, r_{2},\ \zeta
_{a}\rangle=\]\vspace*{-9mm}
\begin{equation}
  =\left(\!\!
\begin{array}
[c]{c}%
if_{0}(r_{2})\Omega_{1/2,1/2,1}^{\prime}(2)\\[2mm]
g_{0}(r_{2})\Omega_{1/2,1/2,0}(2)
\end{array}
\!\!\right)\!\!  ,
\end{equation}\vspace*{-5mm}
%36в
\[ |3\rangle=|2P_{1/2}, +1/2, r_{1}, \zeta_{b}%
\rangle=\]\vspace*{-9mm}
\begin{equation}
  =\left(\!\!
\begin{array}
[c]{c}%
i\tilde{g}_{0}(r_{1})\Omega_{1/2,1/2,1}^{\prime}(1)\\[2mm]
-\tilde{f}_{0}(r_{1})\Omega_{1/2,1/2,0}(1)
\end{array}
\!\!\right)\!\!  ,
\end{equation}\vspace*{-5mm}
%36г
\[
 |4\rangle=\hat{U}_{c}|2P_{1/2}, -1/2, r_{2},
\zeta _{b}\rangle=\]\vspace*{-9mm}
\begin{equation}
  =\left(\!\!
\begin{array}
[c]{c}%
\tilde{f}_{0}(r_{2})\Omega_{1/2,1/2,0}(2)\\[2mm]
-i\tilde{g}_{0}(r_{2})\Omega_{1/2,1/2,1}^{\prime}(2)
\end{array}
\!\!\right)\!\!  ,
\end{equation}\vspace*{-5mm}
%36д
\[
  |5\rangle=|2S_{1/2}, +1/2, r_{1}, \zeta\rangle
=\]\vspace*{-9mm}
\begin{equation}
  =\left(\!\!
\begin{array}
[c]{c}%
g_{1}(r_{1})\Omega_{1/2,1/2,0}(1)\\[2mm]
if_{1}(r_{1})\Omega_{1/2,1/2,1}^{\prime}(1)
\end{array}
\!\!\right)\!\!  ,
\end{equation}\vspace*{-5mm}
%36е
\[ |6\rangle=\hat{U}_{c}|2S_{1/2}, -1/2, r_{2}, \zeta
\rangle=\]\vspace*{-9mm}
\begin{equation}
  =\left(\!\!
\begin{array}
[c]{c}%
i f_{1}(r_{2})\Omega_{1/2,1/2,1}^{\prime}(2)\\[2mm]
g_{1}(r_{2})\Omega_{1/2,1/2,0}(2)
\end{array}
\!\!\right)\!\!  .
\end{equation}
\end{subequations}
Here, $r_{1,2}=\sqrt{(z\pm R/2)^{2}+\rho^{2}}$,
$e^{i\theta_{1,2}}=\frac{z\pm R/2+i\rho}{r_{1,2}}$,
$\rho=\sqrt{x^{2}+y^{2}}$ (see Fig.~4), and the radial functions
$f_{i}(r)$, $g_{i}(r)$, $\tilde{f}_{i}(r)$, and $\tilde{g}_{i}(r)$
are determined by numerically integrating the problem with one
center (see Section \ref{Onenucleus}).

In addition to the true charge of a nucleus, $\zeta$, two effective
charges $\zeta_{a,b}$ are introduced.\,\,They are required in order
to extend the applicability range of our method onto supercritical
nuclear charges.\,\,The
effective charges have the following values:%
%37
\begin{equation}
\begin{array}{l}
\zeta_{a}=\left\{\!\!
\begin{array}{ll}
\zeta, & \zeta<\zeta_{c1}\!,\\[1mm]
0.99\ \zeta_{c1},& \zeta\geq\zeta_{c1}\!,
\end{array}
\right.\\[3mm]
\zeta_{b}=\left\{\!\!
\begin{array}{ll}
\zeta, & \zeta<\zeta_{c2},\\[1mm]
0.99\ \zeta_{c2},& \zeta\geq\zeta_{c2},
\end{array}
\right.
\end{array}
\end{equation}
where $\zeta_{c1}$ and $\zeta_{c2}$ are the critical charge values, at which
the first and second, respectively, energy levels plunge into the lower continuum.

Let us project the Dirac equation $\hat{H}|\Psi\rangle=E|\Psi\rangle$ onto
each of the $|i\rangle$ states ($i=1,...,6$):%
%38
\begin{equation}
\sum\limits_{j=1}^{6}(H_{ij}-ES_{ij})c_{j}=0.
\end{equation}
Then the condition of non-triviality for the coefficients in the linear
combination (\ref{linear-combination2}) gives rise to the secular equation%
%39
\begin{equation}
\det\left\vert H_{ij}-\varepsilon S_{ij}\right\vert =0, \label{secular}%
\end{equation}
where $H_{ij}=\langle i|\hat{H}|j$ are the matrix elements, and $S_{ij}%
=\langle i|j\rangle$ the overlap integrals ($i,j=1,...,6$).

The overlap integrals are calculated in Appendix. Of non-diagonal
integrals, only 4 are independent; these are $S_{14}$, $S_{15}$,
$S_{16}$, and $S_{45}$.\,\,Therefore, the Gram matrix looks like
%40
\begin{equation}
\mathbf{S}=\left(\!\!
\begin{array}{cccccc}
1 & 0 & 0 & S_{14} & S_{15} & S_{16}\\
0 & 1 & -S_{14} & 0 & -S_{16} & S_{15}\\
0 & -S_{14} & 1 & 0 & 0 & -S_{45}\\
S_{14} & 0 & 0 & 1 & S_{45} & 0\\
S_{15} & -S_{16} & 0 & S_{45} & 1 & 0\\
S_{16} & S_{15}  & -S_{45} & 0 & 0 & 1
\end{array}
\!\!\!\right)\!\!.
\end{equation}
%Fig. 7
\begin{figure}
\vskip1mm
  \includegraphics[width=\column]{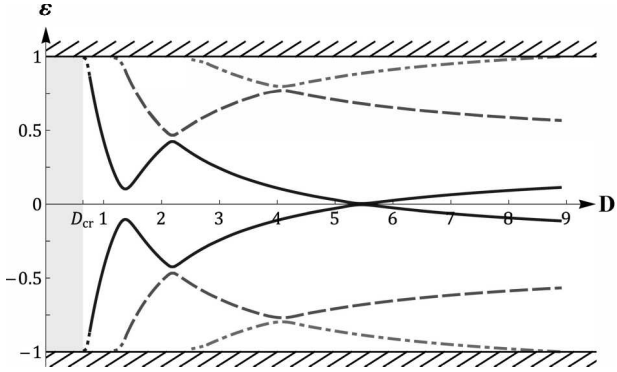}
\vskip-3mm\caption{Dependences of the energy of an electron in the
field of an electric dipole on the dimensionless dipole moment
$D=\frac{D_{\rm ph}}{ea_{0}}=\zeta R$ for the nuclear charge
$Z=174$. Different types of curves correspond to different energy
levels. The tinted area marks the region with no bound states [see
formula (\ref{criticaldipole})]}
  \label{LCAO_energy2}
\end{figure}
%Fig. 8
\begin{figure}
\vskip3mm
  \includegraphics[width=\column]{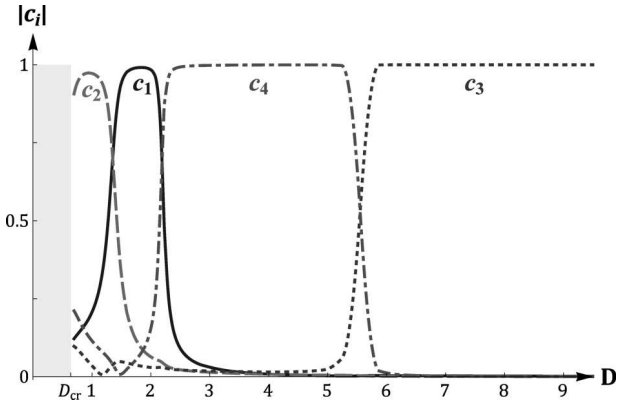}
\vskip-3mm\caption{Dependences of coefficients in the linear
combination (\ref{linear-combination2}) for the highest filled
electron state (the lower
solid curve in Fig.~5) on the dimensionless dipole moment $D=\frac{D_{\rm ph}%
}{ea_{0}}=\zeta R$ for the nuclear charge $Z=174$. See further
explanations in the text }
  \label{LCAO_coef2}
\end{figure}

\noindent
While calculating the matrix elements, it is convenient to present the total
Hamiltonian (\ref{hamiltonian-dipole}) in the following form:%
%41
\begin{equation}
\hat{H}(\zeta)= \left\{\!\!
\begin{array}{l}
\hat{H}_{-}(\zeta_{i},r_{1})+\zeta
v(r_{2})-(\zeta-\zeta_{i})v(r_{1}),\\[2mm]
\hat{H}_{+}(\zeta_{i},r_{2})-\zeta
v(r_{1})+(\zeta-\zeta_{i})v(r_{2}),
\end{array}
\right.
\end{equation}
where $\zeta_{i}=\left\{  \zeta,\ \zeta_{a},\ \zeta_{b}\right\}
$.\,\,The advantage of this representation consists in that the
states $|i\rangle$ are characteristic of $\hat{H}_{\pm}$:
%42
\begin{equation}
\begin{array}{l}
\hat{H}_{-}(\zeta_{a},r_{1})|1\rangle=\varepsilon^{(1S)}(\zeta_{a})|1\rangle,\\[2mm]
\hat{H}_{+}(\zeta_{a},r_{2})|2\rangle=-\varepsilon^{(1S)}(\zeta_{a})|2\rangle,\\[2mm]
\hat{H}_{-}(\zeta_{b},r_{1})|3\rangle=\varepsilon^{(2P)}(\zeta_{b})|3\rangle,\\[2mm]
\hat{H}_{+}(\zeta_{b},r_{2})|4\rangle=-\varepsilon^{(2P)}(\zeta_{b})|4\rangle,\\[2mm]
\hat{H}_{-}(\zeta,r_{1})|5\rangle=\varepsilon^{(2S)}(\zeta)|5\rangle,\\[2mm]
\hat{H}_{+}(\zeta,r_{2})|6\rangle=-\varepsilon^{(2S)}(\zeta)|6\rangle.
\end{array}
\end{equation}
The calculation of matrix elements reveals that there are only
twelve
independent ones:%
%43
\begin{subequations}
\begin{equation}
H_{11}=-H_{22}=\varepsilon^{(1S)}(\zeta_{a})+\zeta
C^{(2)}_{1}-(\zeta-\zeta_{a})C^{(1)}_{1},
\end{equation}\vspace*{-9mm}
\begin{equation}
H_{12}=-(2\zeta-\zeta_{a})A^{(1)}_{12},
\end{equation}\vspace*{-9mm}
\begin{equation}
H_{13}=-H_{24}=\zeta A^{(2)}_{13},
\end{equation}\vspace*{-9mm}
\begin{equation}
H_{14}=H_{23}=-\varepsilon^{(2P)}(\zeta_{b})S_{14}-\zeta
A^{(1)}_{14}+(\zeta-\zeta_{b})A^{(2)}_{14},
\end{equation}\vspace*{-9mm}
\begin{equation}
H_{15}=-H_{26}=\varepsilon^{(2S)}(\zeta)S_{15}+\zeta A^{(2)}_{15},
\end{equation}\vspace*{-9mm}
\begin{equation}
H_{16}=H_{25}=-\varepsilon^{(2S)}(\zeta)S_{16}-\zeta A^{(1)}_{16},
\end{equation}\vspace*{-9mm}
\begin{equation}
H_{33}=-H_{44}=\varepsilon^{(2P)}(\zeta_{b})+\zeta
C^{(2)}_{3}-(\zeta-\zeta_{b})C^{(1)}_{3}\!,
\end{equation}\vspace*{-9mm}
\begin{equation}
H_{34}=-(2\zeta-\zeta_{b})A^{(1)}_{34},
\end{equation}\vspace*{-9mm}
\begin{equation}
H_{35}=-H_{46}=\zeta A^{(2)}_{35},
\end{equation}\vspace*{-9mm}
\begin{equation}
H_{36}=H_{45}=\varepsilon^{(2S)}(\zeta)S_{45}+\zeta A^{(2)}_{45},
\end{equation}\vspace*{-9mm}
\begin{equation}
H_{55}=-H_{66}=\varepsilon^{(2S)}(\zeta)+\zeta C^{(2)}_{5},
\end{equation}\vspace*{-9mm}
\begin{equation}
H_{56}=-\zeta A^{(1)}_{56},
\end{equation}
\end{subequations}
where new notations for the exchange, \mbox{$A_{ij}^{(k)}=$}\linebreak \mbox{$=\langle i|v(r_{k}%
)|j\rangle$} \mbox{($i\neq j$)}, and Coulomb, \mbox{$C_{i}^{(k)}=$} \mbox{$=\langle i|v(r_{k}%
)|i\rangle$}, integrals are introduced.\,\,All in\-de\-pen\-dent
ex\-change and Coulomb integrals are cal\-cu\-la\-ted in Appendix.

The spectrum of the system is found from the secular equation (\ref{secular}), and
the coefficients $c_{i}$ of a linear combination are determined as the
corresponding normalized characteristic vectors:%
%44
\begin{equation}
(\mathbf{H}-\varepsilon\mathbf{S})\vec{c}=0,\ \left(  \vec{c}\right)
^{\dagger}\mathbf{S}\vec{c}=1.
\end{equation}

If the nuclei that compose the dipole approach each other very much,
they become partially screened and effectively reduce their
charges.\,\,As they move apart, this mutual screening
decreases.\,\,Therefore, the growth of the distance between the
nuclei is equivalent, to a certain extent, to a gradual increase of
the charges at the motionless nuclei from zero to real
\mbox{values.}

Figure~7 illustrates the energy dependences for the electron in the
field of an electric dipole in the case of nuclear charge $Z=174$
($\zeta=1.27$).\,\,From Fig.~3, one can see that, at $Z=174$, the
first level has already crossed the whole mass gap and plunged into
the lower continuum.\,\,At the same time, the second level (for the
positive nucleus) has already intersected with the first level for
the negative nucleus and also crossed the zero energy value, so that
$\varepsilon^{(2P)}=-0.28<0$.\,\,Those facts are reflected in the
behavior of the energy and wave functions of the electron in the
dipole potential.

The first \textquotedblleft constriction\textquotedblright\ of solid
curves in Fig.~7 corresponds to the passage of the first energy
level of the electron in the field of one nucleus through the zero
value (this case was considered in the previous section).\,\,As was
shown earlier, the passage of this constriction is accompanied by
the migration of the wave function from the negative nucleus to the
positive one, which is shown in Fig.~8): the coefficient $c_{2}$
dominates before the constriction, and the coefficient $c_{1}$ after
it.\,\,Analogously to what was done in the previous section, this
migration can be interpreted as the creation of an electron-positron
pair from vacuum, with the new particles being in the bound state
with the nuclei.

The relative approach and moving away of the solid and dashed curves
at $D\approx2.2$ in Fig.~7 corresponds to the intersection between
the first level in the field of the positive nucleus and the second
level in the field of the negative nucleus (and {\it vice
versa}).\,\,At this point, an electron transits from the filled
second level of the negatively charged nucleus (the dashed curve) to
the unfilled first level of the same nucleus (the solid
curve).\,\,As a result, the coefficient $c_{4}$ starts to dominate
(see Fig.~8).\,\,Simultaneously, an electron near the positive
nucleus transits from the first level to the empty second level,
thereby vacating the place and creating preconditions for the next
pair to emerge.

Finally, the second constriction in the solid curves in Fig.~7
corresponds to that the second energy level of the electron in the
field of one nucleus crosses the zero value.\,\,The passage of the
constriction is accompanied by the relocalization of the wave
function from the negative nucleus to the positive one (the
coefficient $c_{4}$ dominates before the constriction, and the
coefficient $c_{3}$ after it; see Fig.~8).\,\,Similarly to the
previous case, this is interpreted as the creation of the second
electron-positron pair from \mbox{vacuum.}\looseness=1

Hence, the gradual adiabatic growth of the dipole moment gives rise to the
creation of electron-positron pairs from vacuum owing to the phenomenon of
wave function migration for the \textquotedblleft Dirac sea\textquotedblright%
\ electrons.\,\,Their maximum number is equal to the number of
energy levels of the electron in the field of one nucleus (for a
given value of its charge) that crossed the zero energy value.

\section{Conclusions}\vspace*{1.5mm}

\label{conclusions}

To summarize, a new type of supercritical instability of Dirac
electrons in the potential of a finite electric dipole composed by
two opposite charged nuclei has been studied.\,\,In this geometry,
the Dirac equation does not allow the separation of variables, so
that approximate methods were used.\,\,To calculate the spectrum and
wave functions of an electron in the field of the dipole, the method
of linear combinations of atomic orbitals (LCAO) was
applied.\,\,This method is rather simple for a realization and
allows the majority of calculations to be carried out
analytically.\,\,At the same time, it is inapplicable at short
distances between the impurities.\,\,The wave functions of the
ground state and several lowest excited states of electron in the
potential of one nucleus, which are centered at the corresponding
nucleus, were used as atomic orbitals.

If the charge of each nucleus is large enough for the lowest bound
state of electron in the field of one of the nuclei to cross the
level $E=0$ (i.e.\,\,the electron and positron levels should
together overlap the interval equal to $2mc^{2}$), the gradual
increase of the distance between the center of nuclei results in
that the electron and positron levels first start from the
corresponding continua and approach, by tending to intersect, each
other.\,\,However, in accordance with the Wigner--von Neumann
theorem concerning the absence of a level crossing, they do not
intersect, but start to move apart and asymptotically approach the
levels obtained in the field of one center.\,\,Thus, the spectrum
has a characteristic \textquotedblleft
constriction\textquotedblright.\,\,While passing it, the wave
function of the electron changes its localization.\,\,The electron
in the highest filled state migrates from the negatively charged
center to the positively charged one.\,\,Apparently, it looks as if
the electron generated from the Dirac vacuum becomes localized at
the positive nucleus and screens it, whereas the positron is
localized at the negative nucleus.\,\,In other words, similarly to
the case of supercritical instability with identically charged
nuclei, an electron-positron pair is created; however, now the
particles are created in the bound state.\,\,It is also demonstrated
that the further growth of nuclear charges may result in the
generation of a larger number of electron-positron pairs.

However, if the nuclear charges are too small for the electron and
positron levels, together, to overlap the energy gap $2mc^{2}$, the
described phenomena cannot be observed.\,\,Therefore, the
supercritical instability in the dipole field has a threshold
character.

The threshold values of nuclear charges that are required for the
phenomenon of wave-function migration to be observed are
$Z\sim150$.\,\,Unfortunately, nuclei with so large charges do not
exist in the nature.\,\,Moreover, two nuclei with opposite charge
signs are needed for this purpose.\,\,Those facts make the
experimental observation of this phenomenon practically
impossible.\,\,At the same time, it may hopefully be observed in
graphene, because the threshold charge values are comparable in this
case with unity, and no problems arise concerning the creation of
impurities with opposite charges (positive and negative
ions).\,\,The impurities can be moved using an electron microscope
tip.\,\,If the impurities were first put closer to each other and
afterward moved a large distance apart, they should be
screened.\,\,This is a smoking gun of the described phenomenon.

\vskip3mm

{\it
The\,author\,expresses\,his\,sincere\,gratitude\,to\,E.V.\,Gor\-bar
and V.P.\,Gusynin for their valuable advices and corrections made
during the discussion of this work.\,\,The author also thanks
O.I.\,Voitenko for the qualitative translation of this article from
Ukrainian.}

\subsubsection*{APPENDIX\\
Overlap, Coulomb, and Exchange Integrals}

{\footnotesize In this Appendix, the results of calculations are
presented for the overlap, Coulomb, and exchange integrals that
arise, when the LCAO method is applied to the problems of an electron
in a dipole potential.

Overlap integrals:
%А1,2
\begin{equation}
S_{ii}=1,\ \ S_{ji}=S_{ij}^{*} \ \ \ i,j=\overline{1, 6},\tag{A1}
\end{equation}
\[
S_{12}\equiv S=\int d^{3}r
\left[ig_{0}(r_{1})f_{0}(r_{2})\Omega^{\dagger}(1)\Omega'(2)\,-\right.\]
\[\left.-\,ig_{0}(r_{2})f_{0}(r_{1})\Omega'^{\dagger}(1)\Omega(2)\right]=\]
\[=\frac{1}{2}\!\!\int\limits_{0}^{\infty}\!\! \rho\  d\rho
\!\!\int\limits_{-\infty}^{\infty}\!\!dz\
\left[g_{0}(r_{1})f_{0}(r_{2})
\cos\theta_{2}+g_{0}(r_{2})f_{0}(r_{1})\cos\theta_{1}\right]=
\]
\begin{equation}\label{Sint}
=|\text{in the second term}~ z\rightarrow -z|=0,\tag{A2}
\end{equation}
%A3
\[S_{13}=\!\int\! d^{3}r
\left[i
g_{0}(r_{1})\tilde{g}_{0}(r_{1})\Omega^{\dagger}(1)\Omega'(1)\,+\right.\]
\[\left.+\,if_{0}(r_{1})\tilde{f}_{0}(r_{1})\Omega'^{\dagger}(1)\Omega(1)\right]=\]
\begin{equation}
=\frac{1}{4\pi}\!\int\!
d^{3}r_{1}\left[g_{0}(r_{1})\tilde{g}_{0}(r_{1})-f_{0}(r_{1})\tilde{f}_{0}(r_{1})\right]
\cos\theta_{1}=0,\tag{A3}
\end{equation}
%A4
\[S_{14}=\!\!\int\!\! d^{3}r
\left[
g_{0}(r_{1})\tilde{f}_{0}(r_{2})\Omega^{\dagger}(1)\Omega(2)\,-\right.\]
\[\left.-\,\tilde{g}_{0}(r_{2})f_{0}(r_{1})\Omega'^{\dagger}(1)\Omega'(2)\right]=\]
\[=\frac{1}{2}\!\!\int\limits_{0}^{\infty}\!\! \rho\ d\rho
\!\!\int\limits_{-\infty}^{\infty}\!\!dz\
\left[g_{0}(r_{1})\tilde{f}_{0}(r_{2})-\tilde{g}_{0}(r_{2})f_{0}(r_{1})\cos(\theta_{2}-\theta_{1})\right]=\]
\[=\frac{1}{2}\!\!\int\limits_{0}^{\infty}\!\! \rho\ d\rho
\!\!\int\limits_{-\infty}^{\infty}\!\!dz\
\left[g_{0}(r_{1})\tilde{f}_{0}(r_{2})\,-\right.\]
\begin{equation}
\left.-\,\frac{z^{2}+\rho^{2}-R^{2}/4}{r_{1}r_{2}}\tilde{g}_{0}(r_{2})f_{0}(r_{1})\right]\!\!,\tag{A4}
\end{equation}
%А5
\[S_{15}=\!\int\! d^{3}r
\left[
g_{0}(r_{1})g_{1}(r_{1})\Omega^{\dagger}(1)\Omega(1)\,+\right.\]
\[\left.+\,f_{0}(r_{1})f_{1}(r_{1})\Omega'^{\dagger}(1)\Omega'(1)\right]=\]
\begin{equation}=\!\int\limits_{0}^{\infty}\! dr\ r^{2}
\left[g_{0}(r)g_{1}(r)+f_{0}(r)f_{1}(r)\right]\!.\tag{A5}
\end{equation}
It should be noted that $S_{15}=0$ if $\zeta_{a}=\zeta$, because,
in this case, the functions $|1\rangle$ and $|5\rangle$ are
characteristic functions of the same Hamiltonian,
$\hat{H}_{-}(\zeta)$, but correspond to different energies;
therefore, they are orthogonal to each other.
%A6
\[S_{16}=\!\int\! d^{3}r
\left[i
g_{0}(r_{1})f_{1}(r_{2})\Omega^{\dagger}(1)\Omega'(2)\,-\right.\]
\[\left.-\,ig_{1}(r_{2})f_{0}(r_{1})\Omega'^{\dagger}(1)\Omega(2)\right]=\]
\[=\frac{1}{2}\!\int\limits_{0}^{\infty}\! \rho\  d\rho
\!\int\limits_{-\infty}^{\infty}\!\!dz\
\left[g_{0}(r_{1})f_{1}(r_{2}) \cos\theta_{2}\,+\right.\]
\[\left.+\,g_{1}(r_{2})f_{0}(r_{1})\cos\theta_{1}\right]=|\text{in the second term}~ z\rightarrow
-z|=\]
\begin{equation}
=\frac{1}{2}\!\int\limits_{0}^{\infty}\! \rho\  d\rho
\!\int\limits_{-\infty}^{\infty}\!\!dz\
\left(g_{0}(r_{1})f_{1}(r_{2})-g_{1}(r_{1})f_{0}(r_{2})\right)\cos\theta_{2},\tag{A6}
\end{equation}
%A7
\[S_{23}=\!\!\int\!\! d^{3}r
\left[
\tilde{g}_{0}(r_{1})f_{0}(r_{2})\Omega'^{\dagger}(2)\Omega'(1)\,-\right.\]
\[\left.-\,g_{0}(r_{2})\tilde{f}_{0}(r_{1})\Omega^{\dagger}(2)\Omega(1)\right]=\]
\[=\frac{1}{2}\!\!\int\limits_{0}^{\infty}\!\! \rho\  d\rho
\!\!\int\limits_{-\infty}^{\infty}\!\!dz
\left[\tilde{g}_{0}(r_{1})f_{0}(r_{2})\cos(\theta_{2}-\theta_{1})-g_{0}(r_{2})\tilde{f}_{0}(r_{1})\right]=\]
\[=|z\rightarrow -z|=-\frac{1}{2}\!\!\int\limits_{0}^{\infty}\!\!
\rho\  d\rho \!\!\int\limits_{-\infty}^{\infty}\!\!dz
\bigg[g_{0}(r_{1})\tilde{f}_{0}(r_{2})\,-\]
\begin{equation}
\left.-\,\frac{z^{2}+\rho^{2}-R^{2}/4}{r_{1}r_{2}}\tilde{g}_{0}(r_{2})f_{0}(r_{1})\right]=-S_{14},\tag{A7}
\end{equation}
%A8
\[
S_{24}=\!\int\! d^{3}r \left[-i
g_{0}(r_{2})\tilde{g}_{0}(r_{2})\Omega^{\dagger}(2)\Omega'(2)\,-\right.\]
\[\left.-\,if_{0}(r_{2})\tilde{f}_{0}(r_{2})\Omega'^{\dagger}(2)\Omega(2)\right]=\]
\begin{equation}
=-\frac{1}{4\pi}\!\!\int\!\!
d^{3}r_{2}\!\left[g_{0}(r_{2})\tilde{g}_{0}(r_{2})-f_{0}(r_{2})\tilde{f}_{0}(r_{2})\right]\!\cos\theta_{2}=0,\tag{A8}
\end{equation}
%A9
\[
S_{25}=\!\int\!\! d^{3}r \left[i
g_{0}(r_{2})f_{1}(r_{1})\Omega^{\dagger}(2)\Omega'(1)\,-\right.\]
\[\left.-\,ig_{1}(r_{1})f_{0}(r_{2})\Omega'^{\dagger}(2)\Omega(1)\right]=\]
\[=\frac{1}{2}\!\int\limits_{0}^{\infty}\!\! \rho\  d\rho
\!\int\limits_{-\infty}^{\infty}\!\!dz
\left[g_{0}(r_{2})f_{1}(r_{1}) \cos\theta_{1}\,+\right.\]
\[\left.+\,g_{1}(r_{1})f_{0}(r_{2})\cos\theta_{2}\right]=|\text{in the first term }~ z\rightarrow
-z|=\]
\[
=-\frac{1}{2}\!\!\int\limits_{0}^{\infty}\!\! \rho\ d\rho
\!\!\int\limits_{-\infty}^{\infty}\!\!dz
\Big[g_{0}(r_{1})f_{1}(r_{2})\,-
\]
\begin{equation}
-\,g_{1}(r_{1})f_{0}(r_{2})\Big]\cos\theta_{2}=-S_{16},\tag{A9}
\end{equation}
%A10
\[
S_{26}=\!\int\!\! d^{3}r \left[
g_{0}(r_{2})g_{1}(r_{2})\Omega^{\dagger}(2)\Omega(2)\,+\right.\]
\[\left.+\,f_{0}(r_{2})f_{1}(r_{2})\Omega'^{\dagger}(2)\Omega'(2)\right]=\]
\begin{equation}
=\!\int\limits_{0}^{\infty}\!\! dr\ r^{2}
\left(g_{0}(r)g_{1}(r)+f_{0}(r)f_{1}(r)\right)=S_{15}.\tag{A10}
\end{equation}
%A11
\[
S_{34}=\!\int\!\! d^{3}r \left[-i
\tilde{g}_{0}(r_{1})\tilde{f}_{0}(r_{2})\Omega'^{\dagger}(1)\Omega(2)\,+\right.\]
\[\left.+\,\tilde{g}_{0}(r_{2})\tilde{f}_{0}(r_{1})\Omega^{\dagger}(1)\Omega'(2)\right]=\]
\[=\frac{1}{2}\!\!\int\limits_{0}^{\infty}\!\! \rho\  d\rho
\!\!\int\limits_{-\infty}^{\infty}\!\!dz
\left[\tilde{g}_{0}(r_{1})\tilde{f}_{0}(r_{2})
\cos\theta_{1}+\tilde{g}_{0}(r_{2})\tilde{f}_{0}(r_{1})\cos\theta_{2}\right]=\]
\begin{equation}
=|\text{in the second term}~ z\rightarrow -z|=0,\tag{A11}
\end{equation}
%A12
\[S_{35}=\!\int\!\! d^{3}r
\left[-i
g_{1}(r_{1})\tilde{g}_{0}(r_{1})\Omega'^{\dagger}(1)\Omega(1)\,
-\right.\]
\[\left.-\, if_{1}(r_{1})\tilde{f}_{0}(r_{1})\Omega^{\dagger}(1)\Omega'(1)\right]=\]
\begin{equation}
=\frac{1}{4\pi}\!\!\int\!\!
d^{3}r_{1}\!\left[g_{1}(r_{1})\tilde{g}_{0}(r_{1})-f_{1}(r_{1})\tilde{f}_{0}(r_{1})\right]\!
\cos\theta_{1}=0,\tag{A12}
\end{equation}
%A13
\[
S_{45}=\!\int\!\! d^{3}r \left[
g_{1}(r_{1})\tilde{f}_{0}(r_{2})\Omega^{\dagger}(2)\Omega(1)\,
-\right.\]
\[\left.-\, \tilde{g}_{0}(r_{2})f_{1}(r_{1})\Omega'^{\dagger}(2)\Omega'(1)\right]=\frac{1}{2}\!\!\int\limits_{0}^{\infty}\!\! \rho\ d\rho
\!\!\int\limits_{-\infty}^{\infty}\!\!\!dz
\bigg[g_{1}(r_{1})\tilde{f}_{0}(r_{2})\,-\]
\begin{equation}
-\,\frac{z^{2}+\rho^{2}-R^{2}/4}{r_{1}r_{2}}\tilde{g}_{0}(r_{2})f_{1}(r_{1})\bigg]\!,\tag{A13}
\end{equation}
%A14
\[
S_{36}=\!\int\!\! d^{3}r \left[
-g_{1}(r_{2})\tilde{f}_{0}(r_{1})\Omega^{\dagger}(1)\Omega(2)\,
+\right.\]
\[\left.+\, \tilde{g}_{0}(r_{1})f_{1}(r_{2})\Omega'^{\dagger}(1)\Omega'(2)\right]=\]
\[=-\frac{1}{2}\!\!\int\limits_{0}^{\infty}\!\! \rho\  d\rho
\!\!\int\limits_{-\infty}^{\infty}\!\!dz
\bigg[g_{1}(r_{2})\tilde{f}_{0}(r_{1})\, -\]
\begin{equation}
\left.-\,
\frac{z^{2}+\rho^{2}-R^{2}/4}{r_{1}r_{2}}\tilde{g}_{0}(r_{1})f_{1}(r_{2})\right]=|z\rightarrow
-z|=-S_{45},\tag{A14}
\end{equation}
%A15
\[
S_{46}=\!\int\!\! d^{3}r \left[i
g_{1}(r_{2})\tilde{g}_{0}(r_{2})\Omega'^{\dagger}(2)\Omega(2)\,
+\right.\]
\[\left.+\, if_{1}(r_{2})\tilde{f}_{0}(r_{2})\Omega^{\dagger}(2)\Omega'(2)\right]=\]
\begin{equation}
=-\frac{1}{4\pi}\!\!\int\!\!
d^{3}r_{2}\!\left[g_{1}(r_{2})\tilde{g}_{0}(r_{2})-f_{1}(r_{2})\tilde{f}_{0}(r_{2})\right]\!\cos\theta_{2}=0,\tag{A15}
\end{equation}
%A16
\[S_{56}=\!\int\!\! d^{3}r
\left[i g_{1}(r_{1})f_{1}(r_{2})\Omega^{\dagger}(1)\Omega'(2)\,
-\right.\]
\[\left.-\, ig_{1}(r_{2})f_{1}(r_{1})\Omega'^{\dagger}(1)\Omega(2)\right]=\]
\[=\frac{1}{2}\!\!\int\limits_{0}^{\infty}\!\!\rho\ d\rho
\!\!\int\limits_{-\infty}^{\infty}\!\!dz
\left[g_{1}(r_{1})f_{1}(r_{2})
\cos\theta_{2}+g_{1}(r_{2})f_{1}(r_{1})\cos\theta_{1}\right]=\]
\begin{equation}
=|\text{\rm in the second term }~ z\rightarrow -z|=0,\tag{A16}
\end{equation}

Coulomb integrals:
%A17
\begin{equation}
C^{(1)}_{1}=\!\!\int\limits_{0}^{\infty}\!\!dr\,
r^{2}(f_{0}^{2}(r)+g_{0}^{2}(r))v(r),\tag{A17}
\end{equation}\vspace*{-3mm}
%A18
\begin{equation}
C^{(1)}_{3}=\!\!\int\limits_{0}^{\infty}\!\!dr\,
r^{2}(\tilde{f}_{0}^{2}(r)+\tilde{g}_{0}^{2}(r))v(r),\tag{A18}
\end{equation}\vspace*{-3mm}
%A19
\[
C^{(2)}_{1}\equiv C=\!\!\int\!\! d^{3}r
\frac{1}{4\pi}\left(f_{0}^{2}(r_{1})+g_{0}^{2}(r_{1})\right)v(r_{2})=\]\vspace*{-3mm}
\[=\frac{1}{2}\!\int\limits_{0}^{\infty}\!\!r^{2}dr
\left(f_{0}^{2}(r)+g_{0}^{2}(r)\right)\times\]\vspace*{-1mm}
\[\times\!\!\int\limits_{0}^{\pi}\!\!d\theta\, \sin\theta\, v\!\left(\!\sqrt{r^{2}-2rR
\cos\theta+R^{2}}\right)=\]\vspace*{-1mm}
\[=\frac{1}{2R}\!\int\limits_{0}^{\infty}\!\!dr\,r\left(f_{0}^{2}(r)+g_{0}^{2}(r)\right)\!\!\int\limits_{|r-R|}^{r+R}\!\!dx\,xv(x)=
\]\vspace*{-1mm}
\[
=\frac{1}{R}\!\int\limits_{0}^{R}\!\!dr\,r^{2}\left(f_{0}^{2}(r)+g_{0}^{2}(r)\right)+
\]\vspace*{-1mm}
\begin{equation}
\label{Coul}
+\int\limits_{R}^{\infty}\!\!dr\,r\left(f_{0}^{2}(r)+g_{0}^{2}(r)\right)+\mathcal{O}(r_{0}^{2}),\tag{A19}
\end{equation}\vspace*{-3mm}
%A20
\[C^{(2)}_{3}=\frac{1}{R}\!\!\int\limits_{0}^{R}\!\!dr\,r^{2}(\tilde{f}_{0}^{2}(r)+\tilde{g}_{0}^{2}(r))\,+\]\vspace*{-3mm}
\begin{equation}
+\int\limits_{R}^{\infty}\!\!dr\,r(\tilde{f}_{0}^{2}(r)+\tilde{g}_{0}^{2}(r))+\mathcal{O}(r_{0}^{2}),\tag{A20}
\end{equation}\vspace*{-3mm}
%A21
\[C^{(2)}_{5}=\frac{1}{R}\!\int\limits_{0}^{R}\!\!dr\,r^{2}\left(f_{1}^{2}(r_{1})+g_{1}^{2}(r_{1})\right)+\]\vspace*{-3mm}
\begin{equation}
+\int\limits_{R}^{\infty}\!\!dr\,r\left(f_{1}^{2}(r_{1})+g_{1}^{2}(r_{1})\right)+\mathcal{O}(r_{0}^{2}).\tag{A21}
\end{equation}

Exchange integrals:
%22
\[
 A^{(1)}_{12}\equiv
A=\frac{1}{2}\!\int\limits_{0}^{\infty}\!\!\rho
d\rho\!\!\int\limits_{-\infty}^{\infty}\!\!dz\, g_{0}(r_{1})\,\times
\]
\begin{equation}
\label{Exchange} \times\,f_{0}(r_{2})\frac{z-
R/2}{r_{2}}\left(v(r_{1})-v(r_{2})\right)\!,\tag{A22}
\end{equation}
%A23
\[A^{(2)}_{13}=\frac{1}{2}\!\int\limits_{0}^{\infty}\!\!dr\,r^{2}(g_{0}(r)\tilde{g}_{0}(r)-f_{0}(r)\tilde{f}_{0}(r))\,\times\]\vspace*{-1mm}
\begin{equation}
\times\int\limits_{-1}^{1}\!\!dx\,x\,v\left(\!\sqrt{r^{2}+R^{2}-2Rrx}\right)\!\!,\tag{A23}
\end{equation}
%A24
\[A^{(k)}_{14}=\frac{1}{2}\!\int\limits_{0}^{\infty}\!\!\rho d\rho\!\!\int\limits_{-\infty}^{\infty}\!\!dz\,
\bigg[g_{0}(r_{1})\tilde{f}_{0}(r_{2})\,-\]
\begin{equation}
\left.-\,\frac{z^{2}+\rho^{2}-R^{2}/4}{r_{1}r_{2}}f_{0}(r_{1})\tilde{g}_{0}(r_{2})\right]v(r_{k}),\tag{A24}
\ k=\overline{1,2},
\end{equation}
%A25
\[A^{(2)}_{15}=\frac{1}{2}\!\int\limits_{0}^{\infty}\!\!dr\,r^{2}\left(g_{0}(r)g_{1}(r)+f_{0}(r)f_{1}(r)\right)\times\]
\[\times\int\limits_{-1}^{1}dx\,v\left(\!\sqrt{r^{2}+R^{2}-2Rrx}\right)=\]
\[=\frac{1}{R}\!\int\limits_{0}^{R}\!\!dr\,r^{2}\left(g_{0}(r)g_{1}(r)+f_{0}(r)f_{1}(r)\right)+
\tilde{f}_{0}(r_{2})\,+\]
\begin{equation}
+\int\limits_{R}^{\infty}\!\!dr\,r\left(g_{0}(r)g_{1}(r)+f_{0}(r)f_{1}(r)\right)+\mathcal{O}(r_{0}^{2}),\tag{A25}
\end{equation}
%26
\[
A^{(1)}_{16}=\frac{1}{2}\!\int\limits_{0}^{\infty}\!\!\rho
d\rho\!\!\int\limits_{-\infty}^{\infty}\!\!dz\,
(g_{0}(r_{1})f_{1}(r_{2})\,-
\]
\begin{equation}
-\,f_{0}(r_{2})g_{1}(r_{1}))\cos\theta_{2}\, v(r_{1}),\tag{A26}
\end{equation}
%A27
\[
A^{(1)}_{34}=\frac{1}{2}\!\int\limits_{0}^{\infty}\!\!\rho
d\rho\!\!\int\limits_{-\infty}^{\infty}\!\!dz\,
\tilde{g}_{0}(r_{1})\,\times
\]
\begin{equation}
\times\,\tilde{f}_{0}(r_{2})\frac{z+R/2}{r_{1}}\left(v(r_{1})-v(r_{2})\right)\!,\tag{A27}
\end{equation}
%A28
\[A^{(2)}_{35}=\frac{1}{2}\!\int\limits_{0}^{\infty}\!\!dr\,r^{2}(g_{1}(r)\tilde{g}_{0}(r)-f_{1}(r)\tilde{f}_{0}(r))\,\times\]
\begin{equation}
\times\int\limits_{-1}^{1}\!\!dx\,x\,v\left(\!\sqrt{r^{2}+R^{2}-2Rrx}\right)\!\!,\tag{A28}
\end{equation}
%29
\[A^{(2)}_{45}=\frac{1}{2}\!\int\limits_{0}^{\infty}\!\!\rho d\rho\!\!\int\limits_{-\infty}^{\infty}\!\!dz\,
\biggl[g_{1}(r_{1})\tilde{f}_{0}(r_{2})\,-\]
\begin{equation}
-\,\frac{z^{2}+\rho^{2}-R^{2}/4}{r_{1}r_{2}}f_{1}(r_{1})\tilde{g}_{0}(r_{2})\biggr]v(r_{2}),\tag{A29}
\end{equation}
%A30
\[
A^{(1)}_{56}=\frac{1}{2}\!\int\limits_{0}^{\infty}\!\!\rho
d\rho\!\!\int\limits_{-\infty}^{\infty}\!\!dz\, g_{1}(r_{1})\,\times
\]
\begin{equation}
\times\,f_{1}(r_{2})\frac{z-
R/2}{r_{2}}\left(v(r_{1})-v(r_{2})\right)\!.\tag{A30}
\end{equation}

}

\vspace*{-5mm}
\rezume{%
О.О.\,Соболь}{НАДКРИТИЧНА НЕСТАБІЛЬНІСТЬ\\ ДІРАКІВСЬКИХ ЕЛЕКТРОНІВ У
ПОЛІ ДВОХ\\ ПРОТИЛЕЖНО ЗАРЯДЖЕНИХ ЯДЕР} {У роботі досліджено
рівняння Дірака для електрона в потенціалі скінченного електричного
диполя за допомогою техніки лінійних комбінацій атомних орбіталей
(ЛКАО). Кулонівський потенціал ядер, що утворюють диполь,
регуляризований шляхом врахування їх скінченних розмірів. Показано,
що при перевищенні деякого критичного значення дипольного моменту
спостерігається новий тип надкритичної нестабільності: хвильова
функція найвищого заповненого електронного рівня змінює свою
локалізацію з від'ємно зарядженого ядра на додатно заряджене, що
можна проінтерпретувати як спонтанне народження з вакууму пари
електрона і позитрона, кожен із яких знаходиться у зв'язаному стані
з відповідним ядром і частково його екранує.}

\end{document}